\documentclass[11pt]{article}
\usepackage{pst-all}
\usepackage{pstricks}
\usepackage{pstcol,pst-fill,pst-grad}
\usepackage{amsfonts}
\usepackage{graphicx}
\usepackage{amsmath}

\makeatletter \@addtoreset{equation}{section}

\newcommand{\be}{\begin{equation}}
\newcommand{\ee}{\end{equation}}
\newcommand{\bea}{\begin{eqnarray}}
\newcommand{\eea}{\end{eqnarray}}
\begin{document}
\date{}
\title{
\textbf{    Qubits from  Black Holes in M-theory on K3 Surface   }\\
\textbf{   } }
\author{ Adil Belhaj$^{1}$, Zakariae  Benslimane$^{2}$,  Moulay Brahim Sedra$^{2}$, Antonio Segui$^{3}$
\hspace*{-8pt} \\
\\
{\small $^{1}$LIRST, D\'epartement de Physique,   Facult\'e
Polydisciplinaire, Universit\'e Sultan Moulay Slimane}\\{ \small
B\'eni Mellal, Morocco }
\\ {\small $^{2}$    D\'{e}partement de Physique, LabSIMO,  Facult\'{e}
des Sciences, Universit\'{e} Ibn Tofail }\\{ \small K\'{e}nitra,
Morocco}\\
 {\small $^{3}$    Departamento de F\'{i}sica Te\'{o}rica, Universidad de Zaragoza, E-50
009-Zaragoza, Spain }}  \maketitle

\begin{abstract}
Using M-theory compactification,  we develop a three factor
separation for the scalar submanifold of $N=2$ seven dimensional
 supergravity  associated with  2-cycles of the K3
surface. Concretely, we give  an  interplay between the three scalar
submanifold factors and the extremal black holes obtained from
M2-branes wrapping such 2-cycles. Then, we show that the
corresponding black hole charges are linked to one, two and four
qubit systems.

\end{abstract}


\thispagestyle{empty}

\newpage \setcounter{page}{1} \newpage

\section{Introduction}
Extremal black  holes   have been  extensively investigated in the
context of string theory and related topics including M and F
theories \cite{1,2,3}. These  objects are  obtained from branes
wrapping nontrivial cycles  in the Calabi-Yau manifolds.  The
corresponding scalar moduli  can  be fixed in terms of the black
brane charges using the  attractor  mechanism \cite{4,5,6}.    The
minimum of the effective potential
 generates fixed values of the stringy scalar fields.   This issue  has been also connected   with quantum
information based on  the qubit formalism\cite{7,8}. In fact, a nice
mapping between  eight  charges of  STU black holes  and  three
qubits have been reported  in \cite{9}.  This link has been derived
from type IIB superstring   compactification on a six-torus $T^6$
using string duality.  Alternative methods  based on  graph theory
using  Adinkras  and toric geometry  have been  developed to support
such a correspondence \cite{10,11,12}.  More precisely,  an Adinkra
graphic representation of the extremal black branes constructed from
the
 toroidal  compactification  of type IIA superstring on $T^n$ have been proposed  in \cite{10}.  This
 analysis has been  generalized to  $n$-superqubits.  Concretely, it
has been shown that the number of bosonic and fermionic states
are $\frac{3^n+1}{2}$ and $\frac{3^n-1}{2}$,  associated with odd
and even geometries on the real supermanifold $T^{n|n}$,
respectively .

The aim of this work is to further extend these works by going
beyond the  toroidal compactification  in type II superstrings. More
precisely, we reconsider the study of the moduli space of M-theory
on the K3 surface as a leading example of Calabi-Yan manifolds. This
analysis   offers a new take on the moduli subspace of 2-cycles on
which M2-branes are wrapped to produce seven dimensional black
holes. The focus is on such geometries which in turn can potentially
be interpreted as the moduli space of three classes of black holes. In
particular, consider in some detail such classes, we find
that they are linked to one, two and four qubits systems.

The organization of this paper is as follows. In section 2,  we
develop a new  three factor  separation   for  the scalar
submanifold of $N=2$ seven dimensional   supergravity associated
with 2-cycles of the K3 surface   on which  M-theory is
compactified. The identification of each factor is based on the
appearance of three different real   2-forms of the K3 surface
appearing in the untwisted and  the twisted sectors. In section 3,
we point out the existence of a link between the extremal black hole
charges and qubit systems using the compactification of M-theory on
the K3 surface. The last section is devoted to discussions and open
questions.

\section{ On  the moduli space of black holes in M-theory on the K3 surface}
In this section,  we reconsider the analysis of the moduli space  of
$N = 2$ supergravity in  seven dimensions arising from the M-theory
on the K3 surface \cite{13,130}. The embedding of extremal black  holes
in such a  compactification provides a new factorization scheme for
the scalar submanifolds associated with  real 2-forms of the K3
surface. This decomposition will be used to establish a mapping with
lower dimensional qubits corresponding to   the untwisted and twisted
sectors. We begin by, briefly, recalling that the K3 surface is  a
$2$-dimensional Calabi-Yau manifold  with a  K\"{a}hler structure
permitting the existence of a global nonvanishing holomorphic
$2$-form. Equivalently, it can be defined as a   K\"{a}hler manifold
with a vanishing first Chern class $c_1 = 0$ and  $\textbf{SU}(2)$ Holonomy
group \cite{14}. It is noted that K3 involves a Hodge diagram
playing a crucial role in the determination of the stringy spectrum
in lower dimensions \cite{15}. The string, M, or F, theory
compactification on K3  preserves only half of the
initial supercharges. A close inspection shows that  one  can
construct such a manifold using different ways, including physical
methods based on supersymmetric linear sigma models in two
dimensions. The famous one concerns the orbifold construction which
is given by $T^4$ modulo discrete isometries of $\textbf{SU}(2)$. Concretely,
we take the 4-torus $T^4$ parameterized by four real coordinates
$x_i,\;(i = 1,\ldots,4)$, subject to the following identifications
\begin{equation}
 x_i = x_i + 1.
 \end{equation}
It is convenient to use the  complex coordinates with the following identification constraints
\begin{equation}
 z_i = z_i + 1,\qquad  z_i = z_i + \imath,\qquad  i=1,2.
 \end{equation}
For simplicity reason,  we consider  the $\mathbb{Z}_2$  symmetry
\cite{14}. In this way,   the    orbifold $T^4/\mathbb{Z}_2$  is obtained from
$T^4$ by imposing an extra  $\mathbb{Z}_2$ symmetry acting on the complex
variables $z_i$ as follows
\begin{equation}
 z_i \to  -z_i.
 \end{equation}
To study the cohomology classes of the  K3 surface,  we look for the
forms on $T^4$ which are invariant under  such  a  symmetry. Indeed,
these invariant forms  belonging to   the untwisted sector are given
by
\begin{equation}
1,\; \quad dz_1\wedge dz_2,\; \quad \overline{dz}_1\wedge \overline{dz}_2,\; \quad
\overline{dz}_i\wedge dz_j,\; \quad dz_1\wedge dz_2 \wedge
\overline{dz}_1\wedge \overline{dz}_2.
 \end{equation}
It is recalled that  $h^{p,q}$ denotes the number of  the
holomorphic and the  anti-holomorphic forms of degree $( p,q )$.
These numbers are listed in the following Hodge diagram
\begin{center}
\begin{tabular}{lllll}
&  & $h^{0,0}$ &  &  \\
& $h^{1,0}$ &  & $h^{0,1}$ &  \\
$h^{2,0}$ &  & $h^{1,1}$ &  & $h^{0,2}$ \\
& $h^{2,1}$ &  & $h^{1,2}$ &  \\
&  & $h^{2,2}$ &  &
\end{tabular}=
\begin{tabular}{lllll}
&  & $1$ &  &  \\
& $0$ &  & $0$ &  \\
$1$ &  & $4$ &  & $1$ \\
& $0$ &  & $0$ &  \\
&  & $1$ &  &
\end{tabular}
\end{center}

It has been shown that  there are  16 fixed points which read as
\begin{equation}
(z_1^i,z_2^i)=(0,0), (0,\frac{1}{2}),
(0,\frac{1}{2}\imath),(0,\frac{1}{2}+\frac{1}{2}\imath)\ldots(\frac{1}{2}+\frac{1}{2}\imath,\frac{1}{2}+\frac{1}{2}\imath)
\end{equation}
modifying the Hodge number $h^{1,1}$.  The blowing up of  all 16
fixed points produces the so-called  K3 surface.  Each fixed
point corresponds to a vanishing 2-sphere. \\ Let us explain how to
blow up a singularity. Locally, the orbifold $T^4/Z_2$ looks like
$\mathbb{C}^2/\mathbb{Z}_2$ which is known  by   $A_1$ space described by the
following equation $xy = z^2$  where $x$,  $y$ and $z$ are
$Z_2$-invariant. This can be related to the coordinates of $ \mathbb{C}^2$ as
follows $z = z_1z_2$, $x = z^2_1$,  $y = z^2_ 2$. This local
geometry  involves a singularity at  $ x = y = z = 0$ which can be
replaced by a two dimensional sphere $S^2$  which is
isomorphic to $\mathbb{C}P^1$. The blowing up of 16 fixed points changes
$h^{1,1}=4$ which becomes $h^{1,1}=4+16=20$. The  Hodge diagram,
involving the untwisted and twisted sectors,  takes the following
form
\begin{center}
\begin{tabular}{lllll}
&  & $h^{0,0}$ &  &  \\
& $h^{1,0}$ &  & $h^{0,1}$ &  \\
$h^{2,0}$ &  & $h^{1,1}$ &  & $h^{0,2}$ \\
& $h^{2,1}$ &  & $h^{1,2}$ &  \\
&  & $h^{2,2}$ &  &
\end{tabular} =
\begin{tabular}{lllll}
&  & $1$ &  &  \\
& $0$ &  & $0$ &  \\
$1$ &  & $20$ &  & $1$ \\
& $0$ &  & $0$ &  \\
&  & $1$ &  &
\end{tabular}
\end{center}

 Many geometrical
properties of the compactification on the  K3 surface  can be obtained
from this simple diagram.   The  central horizontal line gives
information on the moduli  of the K3 surface, associated with
2-cycles,  controlled by the   parameters of the complex and the
 K\"{a}hler  deformations. It is recalled that the K3 surface is a
self-mirror manifold. It turns out  that the Hodge diagram carries
not only geometric but also physical  data of the compactified
theories. The relevant physical information, including the black
holes,   can be obtained from  the second  real cohomology $H^2 (K3,
\mathbb{R})$ of the K3 surface which can be  decomposed  as follows
\begin{equation}
H^2 (K3, \mathbb{R})=H^{2,0}(K3, \mathbb{R})\oplus H^{1,1}(K3, \mathbb{R})\oplus H^{0,2}(K3, \mathbb{R}).
\end{equation}
Its dimensions, as a vector space, is
\begin{equation}
dim\;H^2 (K3, \mathbb{R})=dim\; H^{2,0}(K3, \mathbb{R})+ dim\; H^{1,1}(K3, \mathbb{R})+ dim\;
H^{0,2}(K3, \mathbb{R})=22.
\end{equation}
The compactification of M-theory on the K3 surface  gives $N=2$
supergravity  in seven dimensions \cite{13}. It is recalled that, at
low energy limit,  the eleven dimensional  bosonic  massless fields
of M-theory  are
\begin{equation}
 g_{MN}, \;\;C_{MNK},\qquad M,N,K = 0,\ldots,10
\end{equation}
which   are  the metric and  the 3-form gauge potential respectively.
The dimensional reduction on  the K3 surface
 produces the following seven dimensional fields
\begin{equation}
g_{\mu\nu},\;\; C_{\mu\nu\rho},\;\;  C_{\mu ij}, \;\; \phi_a, \qquad
\mu,\nu = 0,\ldots,6, \;\;a=1,\ldots, 58
\end{equation}
where $ g_{\mu\nu}$ is the seven dimensional  metric,  $ C_{\mu ij}$
are  Maxwell gauge fields arising      from the compactification of
$C_{\mu\nu\rho}$  on the real 2-cycles of the K3 surface.  The total
moduli space of  M-theory  on the  K3 surface, parameterized  by
$\phi_a$, reads as
\begin{equation}
\label{modulik3} M= \frac{\textbf{SO}(3,19)}{\textbf{SO}(3)\times \textbf{SO}(19)}\times
SO(1,1).
\end{equation}
In the study of the extremal black objets in M-theory
compactification on the K3 surface,
 four   seven dimensional black  $p$-brane solutions  can appear as
 required  by the  dual condition  \cite{16}
\begin{equation}
p+q=3.
\end{equation}
They are defined by $p=0,1,2,3$ with $AdS_{p+2}\times S^{5-p}$
near-horizon geometries. These solutions  are classified by

\begin{table}[tbph]
\centering
\begin{tabular}{|c|c|c|}
\hline $ p$ & Black object & near-horizon geometry  \\
\hline $ p=0$ & Black holes
& $Ads_{2}\times  S^{5}$ \\ \hline $ p=1$ & Black M-string  &$Ads_{3} \times S^{4} $ \\
\hline $ p=2$& Black M2-brane
& $Ads_{4}\times  S^{3}$ \\ \hline $ p=3$ & Black M3-branes  &$Ads_{5} \times S^{2} $ \\
\hline
\end{tabular}%
\caption{This table gives  the possible  extremal black $p$-brane
solutions in M-theory on K3 surface.}
\end{table}

In fact,
 $p=0$ corresponds to an $AdS_{2}\times S^{5}$ space
 describing the  near-horizon geometry  of  the  electric charged  black
 holes.  They are obtained from M2-branes wrapping 2-cycles in the K3 surface.
 Their magnetic dual  are  black   M3-branes with $AdS_{5}\times S^{2}$ near-horizon
 geometries, which are obtained from M5-branes wrapping also  2-cycles.
The  objects  carry charges associated with the gauge invariant
field strengths $F^i=dA^i(i=1,\ldots,22)$ of the $N=2$ supergravity
theory derived from $C_{\mu\nu\rho}$  on the real 2-cycles of the K3
surface.  However, $p=1$  corresponds to  the $AdS_{3}\times S^{4}$
space describing the  near-horizon geometry  of an extremal
 black string.  It can be  obtained from a  M5-brane wrapping  the K3
 surface. It is  charged under  the   3-form field strengths obtained from the compactification of the K3 surface. The electric charge is
proportional to the integral of the dual of  $H=dC$ over a
3-cycle. The magnetic dual horizon geometry reads as
$AdS_{4}\times S^{3}$ and  it describes  a black M2-brane. The
latter is charged under the gauge invariant 4-form field strengths
$H$.

 Combining the Hodge diagram and the moduli space information of the K3 surface, we can establish a  mapping  between  the black object  charges
and  the moduli space of the  K3 surface. A priori, there are many ways
to approach such  a space depending on addressed questions\cite{16}.
In fact, it is recalled that the moduli
 space ${\cal M}$  can be  associated with the  forms on the K3 surface. Indeed, the first factor $\frac{\textbf{SO}(3,19)}{\textbf{SO}(3)\times \textbf{SO}(19)}$
 is associated with $22$ real 2-forms
\begin{equation}
\frac{\textbf{SO}(3,19)}{\textbf{SO}(3)\times \textbf{SO}(19)} \to \{ \mbox{2-forms} \}.
\end{equation}
However, the remaining  factor corresponds to  the 0-form and its
dual 4-form
\begin{equation}
 \textbf{SO}(1,1) \to \{ \mbox{0-form},\; \mbox{dual 4-form} \}.
\end{equation}
This factorization is directly   related to  the classification of
the black object charges in seven dimensions obtained from M2 and
M5-branes discussed above. In fact, we have the following mapping
\begin{eqnarray}
\frac{\textbf{SO}(3,19)}{\textbf{SO}(3)\times \textbf{SO}(19)} &\to & \{ p=0,\;\; p=3 \}\nonumber \\
{ \textbf{SO}(1,1)} &\to & \{ p=1,\;\; p=2 \}.
\end{eqnarray}

The study of the attractor horizon geometries of extremal black
$p$-branes   on the $K3$ surface   can be explored to get the
complete  correspondence
\begin{table}[tbph]
\centering
\begin{tabular}{|c|c|c|c|}
\hline Coset space & Black brane & Cycles in K3& Gauge symmetry  \\
\hline $\frac{ \textsc{ \textbf{SO}(3,19})}{ \textsc{\textbf{SO}(3)}\times
\textsc{\textbf{SO}(16)}}$ & black holes (black M3-branes)
& 2-cycles & $ \textsc{  \textsc{\textbf{U}(1)}}^{22}$ \\ \hline $ \textsc{\textbf{SO}(1,1)}$ & Black M-string (black M2-brane) & K3 (0-cycle)  &$ \textsc{\textsc{\textbf{U}(1)}}$ \\
\hline
\end{tabular}%
\caption{This table gives  the correspondence between the scalar
manifold factors and  extremal black $p$-brane charges  in M-theory
on  the K3 surface.} \label{table1}
\end{table}

\section{ Embedding qubits in M-theory black holes }
In this section,  we will show that qubits  can be embedded   in
M-theory on the K3 surface.  Concretely, we would like  to elaborate
a link between black holes  and the moduli subspace
$\frac{\textbf{SO}(3,19)}{\textbf{SO}(3)\times \textbf{SO}(19)} $. This correspondence will be
explored to engineer
 qubit systems from the corresponding black holes (black M3-branes).
It is worth   to recall  that the qubit is a primordial building  block
in quantum version of
 information  theory \cite{17,18,19}. This piece  describes  a   system  with two physical states.
 Using Dirac notation, a  single  qubit reads as
\begin{equation}
|\psi\rangle=c_0|0\rangle+c_1 |1\rangle
\end{equation}
 where $c_i$  are complex  numbers   verifying  the  normalization
condition
\begin{equation}
|c_0|^2+|c_1 |^2=1.
\end{equation}
 Geometrically, this probability condition  generates a two-sphere called  Bloch
sphere.  Similarly,  T
the two qubits are four state  systems taking
the following form
\begin{equation}
|\psi\rangle=c_{00}|00\rangle+c_{10}
|10\rangle+c_{01}|01\rangle+c_{11} |11\rangle
\end{equation}
where $c_{ij}$  are complex coefficients  satisfying the
normalization condition
\begin{equation}
|c_{00}|^2+|c_{10}|^2+|c_{01}|^2+|c_{11}|^2=1.
\end{equation}
This equation describes a   three dimensional complex projective
space $\mathbb{C}P^3$ extending  the Bloch sphere. This  analysis can be
generalized  to $n$ qubits associated with  $2^n$  state systems
placed on $\mathbb{C}P^{2^n-1}$   state geometry. An inspection shows  that
there is a direct correspondence between qubits and black  brane
charge vectors in M-theory on the K3 surface. These black solutions,
associated with $p=0$ and $p=3$, can be obtained from M2 and
M5-branes wrapping 2-cycles in the K3 surface respectively.  To get
such a connection,
  we shall reconsider the study  of  the second cohomology  space
$H^2 (K3, \mathbb{R})$ of the K3 surface. In particular, we shall use a
special factorization of the corresponding moduli space.  The
orbifold construction can be  worked out to   show that  the moduli
subspace of M-theory on K3 surface  should have a prior three
factors.
   Under this hypothesis, the  scalar submanifold  $\frac{\textbf{SO}(3,19)}{\textbf{SO}(3)\times \textbf{SO}(19)}$ associated with 2-forms should
take the form
\begin{equation}
\frac{\textbf{SO}(3,19)}{\textbf{SO}(3)\times \textbf{SO}(19)}=M_{1}\times M_{2}\times M_{3}.
\label{fac}
\end{equation}
We will show that this  factorization   can   be related
 to the  existence of three  different classes of  2-cycles  embedded in  the K3 surface.
The identification of each factor can be obtained by the help of the
the second  real cohomology  $H^2 (K3, \mathbb{R})$.  Indeed, the
submanifold group $\frac{\textbf{SO}(3,19)}{\textbf{SO}(3)\times \textbf{SO}(19)}$
    can be factorized as follows
    \begin{eqnarray}
\frac{\textbf{SO}(3,19)}{\textbf{SO}(3)\times \textbf{SO}(19)} = \frac{\textbf{SO}(3,3)}{\textbf{SO}(3)\times
\textbf{SO}(3)} \times  \frac{\textbf{SO}(3,19)}{\textbf{SO}(3)\times \textbf{SO}(19)}.
\end{eqnarray}
The submanifold $\frac{\textbf{SO}(3,3)}{\textbf{SO}(3)\times \textbf{SO}(3)}$  carries
information on the untwisted sector associated with the former   Hodge
diagram. The second factor  corresponds  to  the  twisted sector
controlled by  the 16 fixed points of the orbifold construction.
To embed lower dimensional qubits, we  need to use an extra
factorization. In fact, it is given by
\begin{eqnarray}
\frac{\textbf{SO}(3,19)}{\textbf{SO}(3)\times \textbf{SO}(19)}= \frac{\textbf{SO}(3,2)}{\textbf{SO}(3)\times
\textbf{SO}(2)} \times \frac{\textbf{SO}(3,1)}{\textbf{SO}(3)}\times
\frac{\textbf{SO}(3,16)}{\textbf{SO}(3)\times \textbf{SO}(16)}.
\end{eqnarray}
Then,   we decompose the untwisted sector to
\begin{eqnarray}
 \frac{\textbf{SO}(3,2)}{\textbf{SO}(3)\times
\textbf{SO}(2)} \times \frac{\textbf{SO}(3,1)}{\textbf{SO}(3)}\to\frac{\textbf{SO}(2,2)}{\textbf{SO}(2)\times
\textbf{SO}(2)} \times \frac{\textbf{SO}(2,1)}{\textbf{SO}(2)}.
\end{eqnarray}
 The factor $\frac{\textbf{SO}(2,2)}{\textbf{SO}(2)\times
\textbf{SO}(2)} $ can be linked with the 2-forms $\overline{dz}_i\wedge dz_j$
belonging to the untwisted sector $H^{1,1} (K3, \mathbb{R})$ . However,
 the factor $\frac{SO(2,1)}{SO(2)}$  will be   related  with  the real parts of the complex
holomorphic and antiholomorphic  2-forms  which  are elements of the
cohomology classes  $H^{2,0}(K3, \mathbb{R})$  and $H^{2,0}(K3, \mathbb{R})$ respectively.  The
twisted sector $\frac{\textbf{SO}(3,16)}{\textbf{SO}(3)\times \textbf{SO}(16)}$ should be
factorized  as follows
\begin{eqnarray}
\frac{\textbf{SO}(3,16)}{\textbf{SO}(3)\times \textbf{SO}(16)} = \frac{\textbf{SO}(1,16)}{\textbf{SO}(16)} \times
\frac{\textbf{SO}(2,16)}{\textbf{SO}(2)\times \textbf{SO}(16)}.
\end{eqnarray}
The  factor $\frac{\textbf{SO}(1,16)}{\textbf{SO}(16)} $  can be associated with the
2-cycles  used in the deformation of the 16  fixed points. However,
 the factor $\frac{\textbf{SO}(2,16)}{\textbf{SO}(2)\times \textbf{SO}(16)}$   carries  information of the quaternionic   structure of the K3 surface which has no role in the
 present discussion. This factor will be omitted. Finally, we consider the following decomposition
  \begin{eqnarray}
\frac{\textbf{SO}(3,19)}{\textbf{SO}(3)\times \textbf{SO}(19)} \to \frac{\textbf{SO}(2,1)}{\textbf{SO}(2)}
  \times \frac{\textbf{SO}(2,2)}{\textbf{SO}(2)\times \textbf{SO}(2)} \times \frac{\textbf{SO}(1,16)}{
  \textbf{SO}(16)}.
\end{eqnarray}
Analyzing  these  moduli space factors and  M2-brane charges, the
total abelian gauge group     takes the following form
\begin{equation}
 {\textbf{U}(1)}^{2}\times  {\textbf{U}(1)}^{4}\times  {\textbf{U}(1)}^{16}.
\end{equation}%
This separation of the charges is governed by  the Hodge diagram
encoding the K3 surface forms.  It is  suggested   that ${\textbf{U}(1)}^{2}$
can be associated with the coset space  $\frac{ { \textbf{SO}(2,1)}}{
{\textbf{SO}(2)}}$. The second factor $ {U(1)}^{4}$ is the abelian gauge
symmetry associated with  four  field strength 2-forms
$F^{ij}$($i,j=1,2$).  This  $ {\textbf{U}(1)}^{4}$ gauge symmetry  can be
decomposed as follows
\begin{equation}
 {\textbf{U}(1)}^{4}= {\textbf{U}(1)}^{2}\times  {\textbf{U}(1)}^{2}.
\end{equation}
 This part has   $ { \textbf{SO}(2)}\times  { \textbf{SO}(2)}$
isotropy symmetry.     The corresponding gauge fields are  1-forms
obtained from the the reduction of the M-theory 3-form on 2-cycles
belonging to the untwisted  $H^{1,1}(K3, \mathbb{R})$. The corresponding four
charges can be related to the scalar submanifold factor  $\frac{ {
\textbf{SO}(2,2)}}{ { \textbf{SO}(2)}\times  { \textbf{SO}(2)}}$.  The last gauge symmetry $
{\textbf{U}(1)}^{16}$ corresponds to  strength 2-forms
$F^{i}$($i=1,\ldots,16$) which are
 obtained from the  the reduction of the 3-form gauge potential $C_{MNK}$  on 2-cycles used in
 the deformation of the singular points. They are elements of
the  twisted  sector  $H^{1,1}(K3, R)$. Finally, we propose the
following mapping
\begin{table}[tbph]
\centering
\begin{tabular}{|c|c|c|c|}
\hline Coset space & Black M-brane objects & Gauge symmetry  \\
\hline $\frac{ { \textbf{SO}(2,1)}}{ { \textbf{SO}(2)} } $ & black holes (black
M3-branes)
& $ {\textbf{U}(1)}^{2}$ \\ \hline $\frac{ { \textbf{SO}(2,2)}}{ { \textbf{SO}(2)} \times { \textbf{SO}(2)} }$ & black holes (black M3-branes) & $%
 {\textbf{U}(1)}^{4}$ \\
\hline $\frac{ { \textbf{SO}(1,16)}}{ { \textbf{SO}(16)} } $ & black holes (black  M3-branes) & $ {\textbf{U}(1)}^{16}$\\
\hline
\end{tabular}%
\caption{This table describes the relation between scalar
submanifold factors,  extremal black hole (brane) charges and  the
gauge symmetry  in seven  dimensions.} \label{table2}
\end{table}

 Form the orbifold construction  point of view,  it  is remarked  that
    the seven dimensional gauge field  can  split into three parts
\begin{equation}
A_\mu^a=A_\mu^\alpha \oplus A_\mu^\beta \oplus A_\mu^\gamma
\end{equation}
which are vectors in ${\textbf{SO}(2)}$, ${\textbf{SO}(2,2)}$ and ${\textbf{SO}(16)}$ symmetries
respectively.
 This  decomposition shows that 22 abelian vector fields can
produce
 three classes of  black hole charges depending on the  2-form type   of
the orbifold $\frac{T^4}{\mathbb{Z}_2}$. This classification allows one to
embed  lower dimensional  qubits in such black holes obtained from a
set of M2-branes wrapping  the associated   2-cycles in the K3
surface.

\subsection{One qubit}
  1-qubit
can be embedded in the  black hole  physics in M-theory on the K3
surface which can be  related to  the  scalar submanifold
$\frac{\textbf{SO}(2,1)}{\textbf{SO}(2)}$ associated with $ {\textbf{U}(1)}^{2}$ gauge
symmetry. The modeling of the 1-qubit can be done as in \cite{9},
using the holomorphic and antiholomorphic forms.  The present qubit
representation  will be  related  to the $\frac{T^4}{\mathbb{Z}_2}$
holomorphic and antiholomorphic 2-forms which are elements of the
cohomology classes $H^{2,0}$ and $H^{0,2}$ respectively
\begin{equation}
\Omega=dz_1\wedge dz_2,\quad
\overline{\Omega}=\overline{dz}_1\wedge\overline{dz}_2.
\end{equation}
In order to establish  the correspondence between these 2-forms and
the basis vectors of the one-qubit system,  we  can explore the
scenario proposed in \cite{9}. Indeed, it is recalled that for a
$(p,q)$-form, the Hodge star action reads as
\begin{equation}
(\varphi,\varphi)\frac{\omega^2}{2}=\varphi\wedge\star\overline{\varphi}
\end{equation}
where $\omega=\imath(\overline{dz_1}\wedge
dz_2+\overline{dz_1}\wedge dz_2)$.  It has been verified that for
the orbifold $\frac{T^4}{Z_2}$, the Hodge star action  reduces to
\begin{equation}
(\varphi,\varphi)\frac{\omega^2}{2}=\varphi\wedge\overline{\varphi}.
\end{equation}
Thus, the vector space calculation produces the following relations
\begin{eqnarray}
(\Omega,\Omega)=(\overline{\Omega},\overline{\Omega})=1\\
(\Omega,\overline{\Omega})=(\overline{\Omega},\Omega)=0.
\end{eqnarray}
This shows that $\{\Omega,\overline{\Omega}\}$ forms a two
dimensional basis.  Due to this calculation,  we can take these
vectors as a basis of 1-qubit. The mapping  is given by
\begin{eqnarray}
\Omega &\to& |0\rangle \\\overline{\Omega}&\to& |1\rangle.
\end{eqnarray}
The associated seven dimensional black hole has electric charges
under $ {\textbf{U}(1)}^{2}$ gauge symmetry
\begin{eqnarray}q_0,\quad q_1.
\end{eqnarray}
To get these charges, one uses the following  real 2-forms
\begin{eqnarray}
Re(\Omega)=\omega_0,\qquad Im(\overline{\Omega})=\omega_1.
\end{eqnarray}
Indeed, each  state $|i\rangle$ corresponds to M2-branes
 with charges $q_{i}$ which are  given by
\begin{equation}
q_{i}=\int_{C^2_{i}}F^2,\qquad i=0,1,
 \end{equation}
 where $F^2=q^{\ell}w_{\ell}$ and where  $C^2_{i}$ are  2-cycles  dual
 to 2-forms $\omega_i$.

\subsection{Two  qubits}
The  2-qubit  can be also embedded in the untwisted sector.
Concretely,  we show that  the (1,1)-forms
$\omega_{ij}=\overline{dz_i}\wedge dz_j$,
 ($i,j=1,2)$ can be worked out to form a basis of states describing  2-qubits. To get the corresponding basis of states, we  need first to
establish a symmetric bilinear   scalar product. Indeed, we propose
the following scalar product
\begin{equation}
(\omega_{ij},\omega_{k\ell})=\int_{\frac{T^4}{\mathbb{Z}_2}}\omega_{ij}\wedge \bot \omega_{k\ell}
 \end{equation}
 where  $\bot$  is an operator acting on (1,1)-forms $\omega_{ij}$ as follows
 \begin{equation}
\bot: \omega_{ij} \to \omega_{i+1j+1} \quad (\mbox{mod\;\;2}).
 \end{equation}
 We will see that the orthogonality  relations can  be obtained by  using the normalized volume form on the  K3 surface
\begin{equation}
\int_{\frac{T^4}{\mathbb{Z}_2}}\frac{w^2}{2}=1.
 \end{equation}
Using the above equations, we can show that
\begin{equation}
(\omega_{ij},\omega_{k\ell})=\delta_{ik}\delta_{j\ell}.
 \end{equation}
 In this way, the  basis state of two qubits can be obtained from the following
 mapping (1,1)-forms of the untwisted sector
 \begin{equation}
\omega_{ij} \to   |i-1\;j-1\rangle,\qquad i,j=1,2.
 \end{equation}
Combining these equations, we can write the following relations
 \begin{equation}
(\omega_{ij},\omega_{k\ell})= \langle
i-1j-1|k-1\ell-1\rangle=\delta_{ik}\delta_{j\ell}.
 \end{equation}
 In connection with seven dimensional  black hole charges, each state $|ij\rangle$  corresponds to  M2-branes
 with charges $q^{ij}$ obtained by the following integration
\begin{equation}
q^{ij}=\int_{C^2_{ij}}F^2.
 \end{equation}
 where $F^2=q^{k\ell}w_{k\ell}$.\\
Moreover, It is possible to recover the entangled  states from a   tensor
representation of (1,1)-forms $\omega_{ij}$. The latter  is  given
by
  \begin{equation}
[\omega_{ij}]=\left(%
\begin{array}{cc}
  \omega_{11} & \omega_{12} \\
 \omega_{21} & \omega_{22} \\
\end{array}%
\right).
 \end{equation}
 The trace of this tensor   reads
   \begin{equation}
\mbox{Trace}\; \omega_{ij}=
  \omega_{11}+ \omega_{22}.
 \end{equation}
 Using   the normalization condition, this trace produces the  following  Bell state
\begin{equation}
\sqrt{2}\;\mbox{Trace}\; \omega_{ij} \to |00\rangle+|11\rangle.
\end{equation}
It is recalled that  the antitrace  can be written as
  \begin{equation}
\mbox{Antitrace}\; \omega_{ij}=
  \omega_{21}-\omega_{12}.
 \end{equation}
 Similarly, we get the following  state
\begin{equation}
\sqrt{2}\;\mbox{Antitrace}\; \omega_{ij} \to |10\rangle-|01\rangle.
\end{equation}
The last comment that  we should make on  the above tensor concerns
its  determinant. By replacing the usual product by wedge
multiplication,  it  reads as
 \begin{equation}
  \mbox{Det}\;
\omega_{ij}=
  \omega_{11}\wedge\omega_{22}- \omega_{21}\wedge\omega_{12}.
 \end{equation}
Integrating this form on the K3 surface,   we  get  the following
constraint on   the black holes charges
  \begin{equation}
 q_{11}q_{22}=q_{21}q_{12}.
 \end{equation}
It  seems that there could be a  link with entanglement and Segre
embedding discussed  in connection with conifold geometries  in type
II superstrings \cite{20}. It is  recalled that the Segre
embedding is considered as  a mapping taking products
 of projective Hilbert spaces  as a projective variety. In fact, it is defined as follows
\begin{equation}
\sigma : P^n \times P^m \rightarrow \Sigma_{n,m} \equiv P^{(n+1)(m+1)-1}
 \end{equation}
 where  $P^n$, $P^m$ are projective Hilbert spaces and $\Sigma_{n,m}$ is known as the Segre variety.
We will give a concise description for the physical case of two
fermionic systems, namely the case $m=n=1$ \cite{20}. Higher
dimensional cases are  straightforward.

To establish the link with entanglement, let us consider two
non-interacting fermionic systems. The corresponding  Hilbert space
reads as $\mathbb{C}^2 \otimes \mathbb{C}^2 = \mathbb{C}^4$.  The
normalization constraint implies that the space of states is
$\mathbb{C}P^3$. Each system has a state representation in terms  of
the Bloch sphere
 $S^2=\mathbb{C}P^1$. Thus,  a general state
$|\Psi\rangle$ can  be represented by  $\mathbb{C}P^1 \otimes
\mathbb{C}P^1$ as follows
\begin{equation}
|\Psi\rangle= |\Psi_1\rangle \otimes |\Psi_2\rangle
\end{equation}
where $ |\Psi_1\rangle = \alpha_1|+\rangle + \beta_1|-\rangle$ and
where  $ |\Psi_2\rangle = \alpha_2|+\rangle + \beta_2|-\rangle$.
This gives
 \begin{equation}
|\Psi\rangle = \alpha_1\alpha_2|++\rangle +
\alpha_1\beta_2|+-\rangle +\alpha_2\beta_1|-+\rangle +\beta_1\beta_2
|--\rangle.
\end{equation}
By rearranging the complex amplitudes in a 4-tuple of projective
coordinates  $$(z_1,z_2,z_3,z_0)\equiv
(\alpha_1\alpha_2,\alpha_1\beta_2,\alpha_2\beta_1,\beta_1\beta_2),$$
we get an hypersurface  embedded in  $\mathbb{C}P^3$ given by the
non linear constraint
\begin{equation}
z_1 z_0= z_2 z_3 \qquad \text{or} \qquad \zeta_1= \zeta_2 \zeta_3
\end{equation}
where $(\zeta_1, \zeta_2, \zeta_3) \equiv (z_1/z_0 , z_2/z_0 ,
z_3/z_0)$ are inhomogeneous coordinates on $\mathbb{C}P^3$. This
constraint   transforms the  K\"{a}hler potential of $\mathbb{C}P^3$
\begin{equation}
K= log(1+|\zeta_1|^2+|\zeta_2|^2+|\zeta_3|^2)
\end{equation}
to
\begin{equation}
K= log(1+|\zeta_2|^2)+log(1+|\zeta_3|^2)
\end{equation}
describing the product of Fubini-Study metrics on $\mathbb{C}P^1
\times \mathbb{C}P^1$. Thus, from a superposition of separable
states $\mathbb{C}P^1 \times \mathbb{C}P^1$, we get entangled states
by the Segre embedding into $\mathbb{C}P^3$. It turns out that
$\mathbb{C}P^1 \times \mathbb{C}P^1$ can be considered as a compact
part of a local Calabi-Yau threefold. The manifold, which  has been
considered as a  small resolution of conifold singularity, has been
used  for  computing exact degeneracies of BPS black holes in type
IIA superstring from D-branes \cite{201}. Thinking about the black
hole/ qubit correspondence, it is possible to make contact with  the
work presented here. We hope  to come back to this issue in future.

It is also noted that  the superoperators discussed in
\cite{202} can be related to the 2-qubit associated with the
$\omega_{ij}$ forms. Interpreting the corresponding Hilbert space as
a Liouvile space, a generalization in terms of a product of the K3
surface can be elaborated to discuss quantum  four valued logic
states. We will address  this question in future works.

\subsection{ Four qubits } In this section,
we treat the black hole charges obtained  from the twisted sector.
This can be related with  4-qubit states.  To do so, we consider the
second homology class  associated with the deformation of the fixed
points
 $(z^i_1,z^i_2)$ of the $\mathbb{Z}_2$ symmetry. It is recalled that,
  in the resolved geometry,   each  fixed point is    replaced by  a 2-cycle
\begin{equation}
(z^i_1,z^i_2)\to C^i_2,\qquad i=1,\ldots, 16
 \end{equation}
Up to  a normalization, the 2-cycles are dual to real  2-forms
$\omega_i$ such that
\begin{equation}
\int_{C^i_2}w_j=\delta_{ij}.
 \end{equation}
To connect this with  a 4-qubit system, we first  need   to
construct the
 scalar  product $(w_i,w_j)$  producing the
orthogonality  relations. To establish the corresponding formulae, one may
explore  certain geometric data of the K3 surface. It is worth
noting
 that, in the K3 surface, there  is a symmetric scalar quantity describing the intersection of  2-cycles.  This information  will be
 explored to work out the scalar product that we are after. Generally for  ALE spaces,
 the singularity  deformation consists on replacing a  fixed  point   by a
collection of intersecting  2-cycles according to the  ADE Dynkin
diagrams. Concretely,  the intersection matrix of the  2-cycles used
for the resolution of ADE singularities is, up to some details, the
opposite of the ADE Cartan matrices $K_{ij}$ \cite{21}. This
eventually leads to a correspondence between the ADE roots and
the 2-cycles. More specifically, to each simple root $\alpha_i$ we
associate a single 2-cycle identified with the complex projective
space $\mathbb{CP}^1$. Inspired by the intersection theory of the K3
surface, we can identify $(w_i,w_j)$ with the Cartan matrix
\begin{equation}
(w_i,w_j)\equiv{C^i_2}.{C^j_2}=- K_{ij}.
 \end{equation}
 It is recalled that the intersection matrix of $n$-dimensional
 sphere $S^n$, used in the blowups   of singularities, is symmetric
 for $n$ even and  antisymmetric for $n$ odd. In the case of the K3
 surface, the matrix is symmetric showing that  the inner
 product is symmetric  and bilinear. This describes  a  particular case of
 hermitian inner product used in the quantum mechanics.

Roughly speaking,  the natural of the singularity  of the orbifold
$\frac{T^4}{\mathbb{Z}_2}$  shows  that the associated algebra is
$\oplus_{i=1}^{16}\textbf{su}(2)$ Lie algebra.
 In this way, the
quantity $(w_i,w_j)$ reads as
\begin{equation}
(w_i,w_j)=-2\delta_{ij}.
 \end{equation}
 To complete the analysis, we can use the Wick rotation and  the following  scale transformation
\begin{equation}
w_i\to \frac{\imath}{\sqrt{2}}w_i.
 \end{equation}
 The 4-qubit  basis of states can be obtained by using an appropriate correspondence. Indeed, to get a
  mapping, we associated to each  fixed  point $(z_1,z_2)$
 a state. Using the cartesian coordinates $z_1=x_1+\imath  x_2$  and $z_2=x_3+\imath  x_4$, we can
 propose  the  following mapping
 \begin{equation}
(z_1,z_2)\to |2x_12x_22x_32x_4\rangle.
 \end{equation}
 representing a 2-cycle  belonging to the twisted sector.
 Using these conventions,  the basis states   are given by
\begin{eqnarray}
(0,0)\to |0000\rangle, \quad (0,\frac{1}{2})\to |0010\rangle, \quad
\ldots \quad , \quad (\frac{1}{2}+\frac{1}{2}\imath,\frac{1}{2}+\frac{1}{2}\imath)\to
|1111\rangle.
\end{eqnarray}
 Clearly, we have  16
quantized charges associated with M2-branes wrapping  on
 $C^{z_1,z_2}_2$ associated with the fixed points $( z_1,z_2)$. Using
 a novel notation associated with such fixed points,  these charges
 take the following matrix form
 \begin{equation}
q_{z_1z_2}=
\left(%
\begin{array}{cccc}
  q_{00} & q_{0\frac{1}{2}} & q_{0\frac{1}{2}\imath}  & q_{0\frac{1}{2}+\frac{1}{2}\imath} \\
  q_{\frac{1}{2}0} & q_{\frac{1}{2}\frac{1}{2}}& q_{\frac{1}{2}\frac{1}{2}\imath} & q_{\frac{1}{2}\frac{1}{2}+\frac{1}{2}\imath} \\
  q_{\frac{1}{2}\imath 0} & q_{\frac{1}{2}\imath\frac{1}{2}}
   & q_{\frac{1}{2}\imath\frac{1}{2}\imath} &q_{\frac{1}{2}\imath\frac{1}{2}+\frac{1}{2}\imath} \\
 q_{\frac{1}{2}+\frac{1}{2}\imath 0} & q_{\frac{1}{2}+\frac{1}{2}\imath\frac{1}{2}} &
 q_{\frac{1}{2}+\frac{1}{2}\imath \frac{1}{2}\imath} & q_{\frac{1}{2}+\frac{1}{2}\imath\frac{1}{2}+\frac{1}{2}\imath} \\
\end{array}%
\right).
 \end{equation}
They  can be computed  by the following integration
\begin{equation}
q_{z_1z_2}=\int_{C^{z_1z_2}_{2}}F^2,
 \end{equation}
 where $F^2$ can be decomposed in terms of  the corresponding 2-forms $
 w_{z_1z_2}$.\\
 A further check shows that there  are many geometric   transformations acting of these
 states which may play the role of gates considered as qubit operators. In particular,  a new
$\mathbb{Z}_2$  symmetry can be implemented  which acts as follows
\begin{equation}
 x \to  x + \frac{1}{2}.
 \end{equation}
This transformation is quite different to  the one  used before. It
has no fixed points associated with  M2 and M5-brane configurations producing new black hole charges.
It is found that there are  15 operations
\begin{equation}
\sum_{k=1}^4C_4^k=15.
 \end{equation}
 For instance,  the operation acting as
 \begin{equation}
 x_i \to  x_i + \frac{1}{2}, \quad i=1,\ldots,4
 \end{equation}
 produces the following 4-qubit operator
\begin{equation}
 |2x_12x_22x_32x_4\rangle \to|2x_1+1\;2x_2+1\;2x_3+1\;2x_4+1\rangle\quad (\mbox{mod\;\;2}).
 \end{equation}
 It has been checked that this operation  can be understood as a 4-flip bit.
\section{Conclusions and open questions}
We have  reconsidered the analysis of the moduli subspace  of $N=2$
seven  dimensional  supergravity obtained from M-theory on  the K3
surface.  Inspired by the Hodge diagram of the K3 surface, we have
proposed
 a new three factor realization for the scalar
submanifold of   $N=2$ supergravity  associated with  the 2-cycles
in the M-theory compactifications. The decomposition  is based on
the existence of three different 2-forms of the K3 surface. In
particular, we have pointed out a correspondence between the scalar
submanifold factors and the extremal black  hole  charges. We  have
specially focused on these factors and their relations to  qubit
systems. More precisely,  we have considered   in some details such
black holes classes, and we  have found that they are linked to one,
two and four qubits systems.

It is observed that the analysis presented here might be
extended to  the case of F-theory on the K3 surface. This
compactification provides an eight dimensional gauge theory
\cite{210,211,212,213}. The F-theory construction requires that the
K3 surface should involve an elliptic fibration structure. It is
known that  its moduli space reads as
\begin{equation}
\label{modulik3} M= \frac{\textbf{SO}(2,18)}{\textbf{SO}(2)\times \textbf{SO}(18)}\times
\mathbb{R}^+\times \mathbb{R}^+
\end{equation}
describing 18 complex parameters and two real ones corresponding to
the K\"{a}hler class of the fiber and the base.

In fact, the elliptic fibration reduces  $\textbf{SO}(3,19)$  to $\textbf{SO}(2,18)$.
Inspired by F-theory/  heterotic string duality in eight dimensions,
the factor $\frac{\textbf{SO}(2,18)}{\textbf{SO}(2)\times \textbf{SO}(18)}$  can be decomposed
\begin{eqnarray}
\frac{\textbf{SO}(2,18)}{\textbf{SO}(2)\times \textbf{SO}(18)}= \frac{\textbf{SO}(2,2)}{\textbf{SO}(2)\times
\textbf{SO}(2)} \times \frac{\textbf{SO}(2,16)}{\textbf{SO}(2)\times \textbf{SO}(16)}.
\end{eqnarray}
Similarity like in M-theory,  the the factor
$\frac{\textbf{SO}(2,16)}{\textbf{SO}(2)\times \textbf{SO}(16)}$ can be reduced to  two copies
of $\frac{\textbf{SO}(1,16)}{ \textbf{SO}(16)}$. Using the equivalence $ \frac{
\textsc{ \textbf{SO}(2,2)}}{\textsc{\textbf{SO}(2)}\times\textsc{\textbf{SO}(2)}}\sim \frac{
\textsc{ \textbf{SO}(2,1)}}{ \textsc{ \textbf{SO}(2)}}\times \frac{ \textsc{
\textbf{SO}(2,1)}}{ \textsc{ \textbf{SO}(2)}}$, the $\frac{\textbf{SO}(2,18)}{\textbf{SO}(2)\times
\textbf{SO}(18)}$  can be decomposed
\begin{eqnarray}
\frac{\textbf{SO}(2,18)}{\textbf{SO}(2)\times \textbf{SO}(18)} \to   \frac{ \textsc{ \textbf{SO}(2,1)}}{
\textsc{ \textbf{SO}(2)}}\times \frac{ \textsc{ \textbf{SO}(2,1)}}{ \textsc{ \textbf{SO}(2)}}
\times \frac{\textbf{SO}(1,16)}{\textbf{SO}(16)}.
\end{eqnarray}
An inspection shows that the factor $ \frac{ \textsc{ \textbf{SO}(2,1)}}{
\textsc{ \textbf{SO}(2)}}\times \frac{ \textsc{ \textbf{SO}(2,1)}}{ \textsc{ \textbf{SO}(2)}}$
correspond  to \textbf{U}(1)$^4$ gauge symmetry. From  heterotic  string
point  of view,  these gauge fields  can be classified in two
categories. Indeed,   U(1)$^2$ are obtained from  the  ten
dimensional metric $g_{\mu\nu}$, while the remaining ones come from
the $B_{\mu\nu}$ field. As in M-theory compactification, $ \frac{
\textsc{ SO(2,1)}}{ \textsc{ SO(2)}}\times \frac{ \textsc{
SO(2,1)}}{ \textsc{ SO(2)}}$ could be linked with two copies of
1-qubit. The last factor \textbf{SO}(1,16)  corresponds  \textbf{U}(1)$^{16}$ gauge
field associated  with Cartan subalgebras of $E_8\times E_8$ (or
\textbf{SO}(32)) gauge symmetry.  We expect  that this factor  can be related
to  a 4-qubit system. We believe  that this connection deserves a
deeper investigation.

Moreover, it will  be interesting to consider the corresponding
superqubit systems. This could be associated with supermanifold
$T^{4,k}$ equipped with four bosonic coordinates and $k$ fermionic
coordinates.  It is recalled that supermanifolds   have been
investigated  in connection with  quantum superlogics\cite{214}. A
possible link can be done by fixing the value of $k$. For case of
$k=4$, this  can produce a K3 supersurface considered as a leading
example of Calabi-Yau supermanifolds. The even and odd submanifolds
can be associated with bosonic and fermionic invariant forms.

 This work comes up with
many open questions. One of them concerns connections with higher
dimensional Calabi-Yau manifolds. It could also be
interesting to look for qubit solutions from  the string Calabi-Yau
moduli space controlled by complex  and K\"{a}hler deformations.
Moreover, many concepts, which have been developed in quantum
information
 theory including gates, circuits and entanglement, could have  geometric representations.  It
should  be of interest to study these issues in the context of the Hodge diagram analysis
 and physical theories related to black  holes and
branes. These issues will be addressed elsewhere.

\textbf{Acknowledgments.}  AS is supported by FPA2012-35453.


\begin{thebibliography}{99}
\bibitem{1}  H. Ooguri, A. Strominger, C.  Vafa, {\em
    Black Hole Attractors and the Topological String},  Phys.Rev.{\bf D70}(2004)106007, {\tt arXiv:hep-th/0405146.
    }
\bibitem{2}  M. Guica, A. Strominger, {\em
    Wrapped M2/M5 Duality },  {\tt arXiv:hep-th/0701011}.


\bibitem{3}  B. Haghighat, S. Murthy, C. Vafa, S. Vandoren, {\em
F-Theory, Spinning Black Holes and Multistring Branches}, {\tt
arXiv:1509.00455}.


\bibitem{4} S. Ferrara, R. Kallosh, A. Strominger, {\em  N = 2 Extremal Black Holes}, Phys. Rev. {\bf
D52}
(1995) 5412,  {\tt hep-th/9508072}.

\bibitem{5} S. Ferrara and R. Kallosh, Supersymmetry and Attractors, Phys. Rev. {\bf D54} (1996) 1514,
{\tt hep-th/9602136}.

\bibitem{6}  R. Ahl Laamara, M. Asorey, A. Belhaj, A, Segui,  {\em
Extremal Black Brane Attractors on The Elliptic Curve}, J.Phys. {\bf A43}
(2010) 105401, {\tt arXiv:0907.0093}.


\bibitem{7} L. Borsten, M. J. Duff, P. L\'evay, {\em
 The black-hole/qubit correspondence: an
up-to-date review},  {\tt arXiv:1206.3166}.
\bibitem{8} L. Borsten, M.J. Duff, A. Marrani, W. Rubens, {\em On the
Black-Hole/Qubit Correspondence},
  Eur.Phys.J.Plus {\bf 126} (2011)
37, {\tt arXiv:1101.3559}.

\bibitem{9} P. L\'evay, {\em  Qubits from extra dimensions},  Phys. Rev. {\bf D
84} (2001) 125020.
\bibitem{10}
    A. Belhaj, M. B. Sedra, A. Segui, {\em   Graph Theory and Qubit Information Systems of Extremal Black
    Branes},  J.Phys. {\bf A48} (2015)  045401,  {\tt arXiv:1406.2578}
\bibitem{11}
    Y. Aadel, A. Belhaj, Z. Benslimane, M. B. Sedra, A. Segui, {\em Qubits from Adinkra Graph Theory via Colored Toric
    Geometry},  {\tt arXiv:1506.02523}.
\bibitem{12}
   A. Belhaj, H. Ez-Zahraouy, M.  B. Sedra,  {\em  Toric Geometry and String Theory Descriptions of Qudit
   Systems}, J.Geom.Phys. {\bf 95} (2015) 21,  {\tt arXiv:1408.3952}.
\bibitem{13} M. B. Schulz, E.  F. Tammaro, {\em
    M-theory/type IIA duality and K3 in the Gibbons-Hawking approximation},  {\tt
    arXiv:1206.1070}.
\bibitem{130}
    T. Okazaki, {\em
Membrane quantum mechanics},  Nucl.Phys. {\bf B890} (2014) 400.


\bibitem{14}  P.  S. Aspinwall, {\em
    K3 Surfaces and String Duality}, {\tt arXiv:hep-th/9611137}.

\bibitem{15}
    C. Vafa, {\em  Lectures on Strings and Dualities}, {\tt
    arXiv:hep-th/9702201}.

\bibitem{16} E. H. Saidi, A. Segui, {\em  Entropy of Pairs of Dual Attractors in six and seven
Dimensions}, {\tt  arXiv:0803.2945 [hep-th]}.
\bibitem{17} M. A. Nielsen,
I. L. Chuang, {\em Quantum Computation and Quantum Information}  Cambridge University Press, New York, NY, USA, 2000.

\bibitem{18} D. R. Terno, {\em
Introduction to relativistic quantum information},  {\tt
arXiv:quant-ph/0508049}.
\bibitem{19}
 M. Kargarian, {\em  Entanglement properties of
topological color codes}, Phys. Rev. {\bf A78} (2008)062312, {\tt
arXiv:0809.4276}.


\bibitem{20} M. Cvetic, G.W. Gibbons, C.N. Pope, {\em  Compactifications of Deformed Conifolds, Branes and the Geometry of
    Qubits}, {\tt arXiv:1507.07585}.



\bibitem{201}

    M. Aganagic, D. Jafferis, N. Saulina, {\em   Branes, Black Holes and Topological Strings on Toric Calabi-Yau Manifolds},
    JHEP{\bf 0612} (2006) 018, {\tt  arXiv:hep-th/0512245}.
\bibitem{202}
  Vasily E. Tarasov,   {\em Quantum Computer with Mixed States and Four-Valued Logic
 Journal of Physics A}. {\bf 25} (2002)
5207, {\tt quant-ph/0312131}.

\bibitem{21}
    S. Katz, P. Mayr, C. Vafa, {\em  Mirror symmetry and Exact Solution of 4D N=2 Gauge Theories I},
    Adv.Theor.Math.Phys.{\bf} 15(1998)53, {\tt
    arXiv:hep-th/9706110}.
\bibitem{210}
C. Vafa, {\em Evidence for F-theory}, Nucl. Phys. {\bf B 469} (1996)
403, {\tt hep-th/9602022}.
\bibitem{211}

V. K. Oikonomou, {\em F-theory Yukawa Couplings and Supersymmetric
Quantum Mechanics}, Nucl.Phys. {\em B856} (2011) 1, {\tt
arXiv:1107.0497}.


\bibitem{212}
V.K.Oikonomou, {\em F-theory and the Witten Index},  Nucl. Phys.
{\bf B850}(2011)273,  {\tt arXiv:1103.1289}.
\bibitem{213}
V. K. Oikonomou, {\em  Graded Geometric Structures Underlying
F-Theory Related Defect Theories},  {\tt arXiv:1303.2537}.
\bibitem{214}
J. Kouneiher, N. Da Costa,  {\em Superlogic Manifolds and Geometric
approach to Quantum Logic}, {\tt  arXiv:1505.00756}.




































\end{thebibliography}
\end{document}